**Glancing-Incidence Focussed Ion Beam Milling: A Coherent X-ray Diffraction Study of 3D Nano-scale Lattice Strains and Crystal Defects**

Felix Hofmann[a], Ross J. Harder[b], Wenjun Liu[b], Yuzi Liu[c], Ian K Robinson[d], Yevhen Zayachuk[e]

[a] Department of Engineering Science, University of Oxford, Parks Road, Oxford, OX1 3PJ, UK
[b] Advanced Photon Source, Argonne National Laboratory, 9700 South Cass Avenue, Argonne, IL 60439 USA
[c] Center for Nanoscale Materials, Argonne National Laboratory, 9700 South Cass Avenue, Argonne, IL 60439 USA
[d] Brookhaven National Laboratory, 734 Brookhaven Avenue, Upton, NY, 11973, USA
[e] Department of Materials, University of Oxford, Parks Road, Oxford, OX1 3PH, UK

*Corresponding Author: felix.hofmann@eng.ox.ac.uk       tel.: +44 1865 283446*

**Abstract:**
This study presents a detailed examination of the lattice distortions introduced by glancing incidence Focussed Ion Beam (FIB) milling. Using non-destructive multi-reflection Bragg coherent X-ray diffraction we probe damage formation in an initially pristine gold micro-crystal following several stages of FIB milling. These experiments allow access to the full lattice strain tensor in the micro-crystal with ~25 nm 3D spatial resolution, enabling a nano-scale analysis of residual lattice strains and defects formed. Our results show that 30 keV glancing incidence milling produces fewer large defects than normal incidence milling at the same energy. However the resulting residual lattice strains have similar magnitude and extend up to ~50 nm into the sample. At the edges of the milled surface, where the ion-beam tails impact the sample at near-normal incidence, large dislocation loops with a range of burgers vectors are formed. Further glancing incidence FIB polishing with 5 keV ion energy removes these dislocation loops and reduces the lattice strains caused by higher energy FIB milling. However, even at the lower ion energy, damage-induced lattice strains are present within a ~20 nm thick surface layer. These results highlight the need for careful consideration and management of FIB damage. They also show that low-energy FIB-milling is an effective tool for removing FIB-milling induced lattice strains. This is important for the preparation of micro-mechanical test specimens and strain microscopy samples.

**Keywords:**
Focussed ion beam; Implantation damage; Coherent X-ray diffraction; Dislocations; Residual lattice strains;

**1. Introduction:**
Focussed ion beam (FIB) techniques have emerged as transformative tools for material analysis and manufacturing at the nano-scale. FIB uses a tightly focussed beam of ions, most commonly gallium (Ga), for nano-scale material imaging, material removal by localised sputtering and deposition by dissociation of precursor gases [1,2]. This ability to view and manipulate materials at the nano-scale has lead to extensive use of FIB for preparation of site-specific microscopy samples, enabling the examination of particular microstructural features by transmission electron microscopy (TEM) [3–6] or atom probe tomography (APT) [7–11]. Combining automated serial FIB sectioning with scanning electron microscopy (SEM) allows the reconstruction of 3D nano-scale sample morphology [12–15] and grain microstructure [16–18].

FIB machining revolutionised the study of nano-scale mechanical properties by enabling the manufacture of micron-sized test specimens. By deforming FIB-machined micro-pillars in a nano-indenter, the dependence of material properties on sample size has been extensively studied, leading to the 'smaller is stronger' paradigm [19–21]. The use of micron-sized mechanical test specimens also makes it possible to examine the properties of materials only available in small volumes. For example FIB-machined micro-cantilevers make it possible to study the mechanical properties of few-micron-thick ion-implanted layers used to simulate irradiation-induced degradation [22,23]. Similarly specimens can be extracted to examine the mechanical properties of specific micro-structural features, such as grain boundaries [24–28], hydrides and phase boundaries [29].

An interesting question concerns the role of damage introduced near the material surface during the FIB milling process. TEM studies have shown that FIB can lead to amorphisation [30], the generation of lattice defects [31,32], formation of intermetallic phases [33], as well as local recrystallisation [34]. When examining the effect of this damage on the behaviour observed in micro-mechanical tests, several competing mechanisms must be considered. FIB-induced crystallographic defects, such as small dislocation loops, act as sources for glide dislocations. This is important when sample strength is controlled by source starvation. Here FIB is expected to cause a substantial reduction in yield stress, as observed in Mo-alloy micro-pillars [35]. This effect is also seen in nano-indentation, where the absence of pop-ins in FIB milled material confirms a greater density of dislocation sources [36]. On the other hand the injection of gallium and the formation of a dense population of defects, are expected to lead to the formation of a hardened surface layer. This would increase the strength of FIB-milled samples [31], and has also been observed in nano-indentation of FIB-exposed molybdenum [36]. The influence of FIB damage on deformation behaviour is reduced in materials with an initially high defect density [35,37]. However, recent measurements, comparing as-FIB-milled Al nano-pillars with pillars that were annealed to remove FIB damage, showed that even though samples from both preparation routes showed similar yield stress, the underlying deformation mechanisms differ markedly [38]. Indeed a strong dependence of mechanical properties of FIB-milled pillars on the exact FIB milling conditions used has been reported [39].

Residual lattice strains introduced by FIB milling play a key role in determining behaviour, as the local stress state controls emission and propagation of glide dislocations. Previous studies of FIB damage showed that stresses of several 100 MPa can be reached within and near the ion-damaged surface layer [40,41]. Indeed in magnesium FIB-induced micro-stresses have been shown to lead to the nucleation of twin domains that extend several microns into the material [42]. Quantifying FIB-induced residual lattice strains and stresses has proven challenging. Thus far the macroscopic deformation of cantilevers [40,43], membranes [44] and thin films [45] has been used to infer FIB-induced lattice strains.

However these rather coarse measurements cannot capture the details of the highly heterogeneous lattice strain fields produced by FIB.

Recently we proposed an alternative approach, using coherent X-ray diffraction to study FIB-induced damage in initially pristine gold micro-crystals [46]. Using this technique we could resolve the full lattice displacement field and hence strain tensor in specific micro-crystals with 10s of nm 3D spatial resolution. Our results showed that even a single FIB imaging scan with low ion dose causes large lattice strains. FIB milling at normal incidence produced an extended network of dislocation loops, which could be mapped out in 3D. In an extensively machined crystal lattice strains extended several 100 nanometers into the crystal, far beyond the ion-implanted layer [41]. This initial study concentrated on studying the effects of normal incidence FIB milling. Yet in the preparation of micro-mechanics test samples, glancing incidence milling is generally used as it provides a better surface finish and, in silicon, has been reported to introduce less damage [34,39,47].

Several approaches have been proposed for the reduction of FIB damage. Most of these use a cleaning step to remove the FIB damaged layer by low energy ion milling, for example using 2 keV Ga ions in the FIB [30] or a low energy Argon ion beam [48]. For silicon, electron-beam-induced-etching with molecular chlorine has also been suggested, though this produced a rough surface finish [49]. An alternative approach, rather than removing the damaged layer, is to anneal samples after FIB milling [38]. This will remove defects, but not the implanted Ga. Furthermore the annealing process is not selective and so may modify defects and microstructure present before FIB milling as well as FIB-induced damage.

In this study we use coherent X-ray diffraction to probe the 3D, nano-scale residual lattice strains and defects produced by glancing incidence FIB machining. By comparing the results to our previous measurements of normal incidence FIB milling damage [46], we examine whether glancing incidence milling indeed produces smaller residual lattice strains. Using 5 keV glancing incidence FIB milling the effectiveness of low energy ion polishing for removing damage from previous FIB processes, and the associated residual lattice strains, is tested. Importantly all our measurements are carried out on the same micro-crystal, imaged at different FIB milling stages, such that a direct comparison can be made. Below we first present details of sample preparation and the coherent X-ray diffraction measurements. This is followed by a discussion of the experimental results and brief conclusions.

**2. Experimental Measurements**
**2.1 Sample Preparation**
Samples were prepared on a [001]-oriented silicon wafer substrate with a 100 nm thick thermally grown oxide layer. The wafer was spin-coated with ZEP resist and electron-beam patterned with a variety of structures, including 2 µm wide lines. After removal of the exposed resist, the wafer was coated with a 3 nm thick Cr adhesion layer followed by 40 nm of Au. Un-patterned areas were removed using a "lift-off" procedure to retain the Au lines. The sample was then annealed at 1273 K in air for 10 hours, after which the gold lines had dewetted to form arrays of gold nano-crystals, ranging in size from ~200 nm to ~1 µm (Fig. 1(a)). Scanning electron microscopy (SEM) was used to inspect the crystals and identify a suitable candidate for this study, avoiding crystals with twin domains, which add complexity and are not of interest here [50].

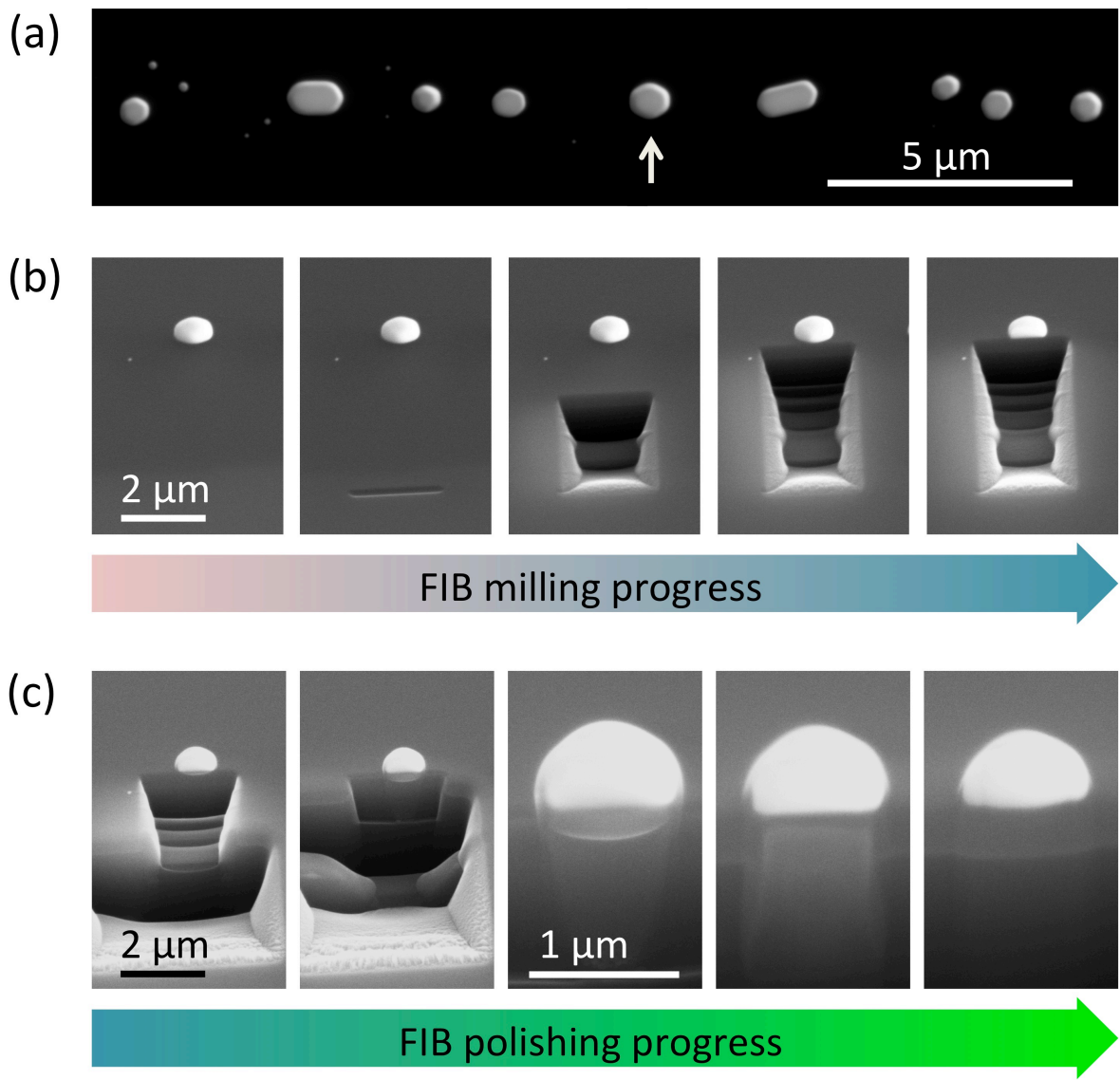

**Fig1: Overview of the sample.** (a) SEM micrograph of the as-made sample, showing a row of gold nano-crystals after dewetting. The crystal under study is highlighted by an arrow. (b) Sequence of SEM images recorded during the 30 keV glancing incidence FIB milling step. (c) Progression of the 5 keV glancing incidence FIB polishing step.

### 2.2 Focussed Ion Beam Milling

Focussed Ion Beam (FIB) milling was carried out on an FEI Nova 600 NanoLab FIB/SEM at the Centre for Nanoscale Materials, Argonne National Lab, USA. Two FIB processing steps were applied, hereafter referred to as 'milling' and 'polishing'.

For the first FIB milling step a Ga ion energy of 30 keV and beam current of 28 pA were used. These conditions closely match those used for the final milling cut during the manufacture of micro-mechanics test specimens [24,51–53]. FIB milling was carried out in an incremental trench-milling mode with a target fluence of 0.65 nC/µm² (4.06 x 10⁹ ions/ µm²) to achieve a glancing-incidence milling condition. Fig. 1 (b) shows SEM images collected during the FIB milling process. Once approximately a third of the crystal had been removed the FIB milling was stopped.

To explore the effectiveness of low energy, glancing incidence FIB milling for the removal of damage introduced by high energy FIB milling, a polishing step with 5 keV ion energy and 150 pA beam current was performed. Again an incremental trench milling mode was used, this time with a target fluence of 3.42 nC/µm² (2.13 x 10¹⁰ ions/ µm²). Fig. 1 (c) shows SEM images collected during the FIB polishing process, which was stopped once a further ~100 nm had been removed from the crystal.

No overview FIB imaging scans were collected to align the FIB processing steps. Instead spatial alignment of FIB and SEM beams was carried out far from the micro-crystal. Then SEM imaging alone was used to position the FIB milling and polishing scans. This is important as our previous results showed that even a single, low dose FIB imaging scan (30 keV, 50 pA, 4.2 x 10⁴ ions/ µm²) can cause large lattice strains [46].

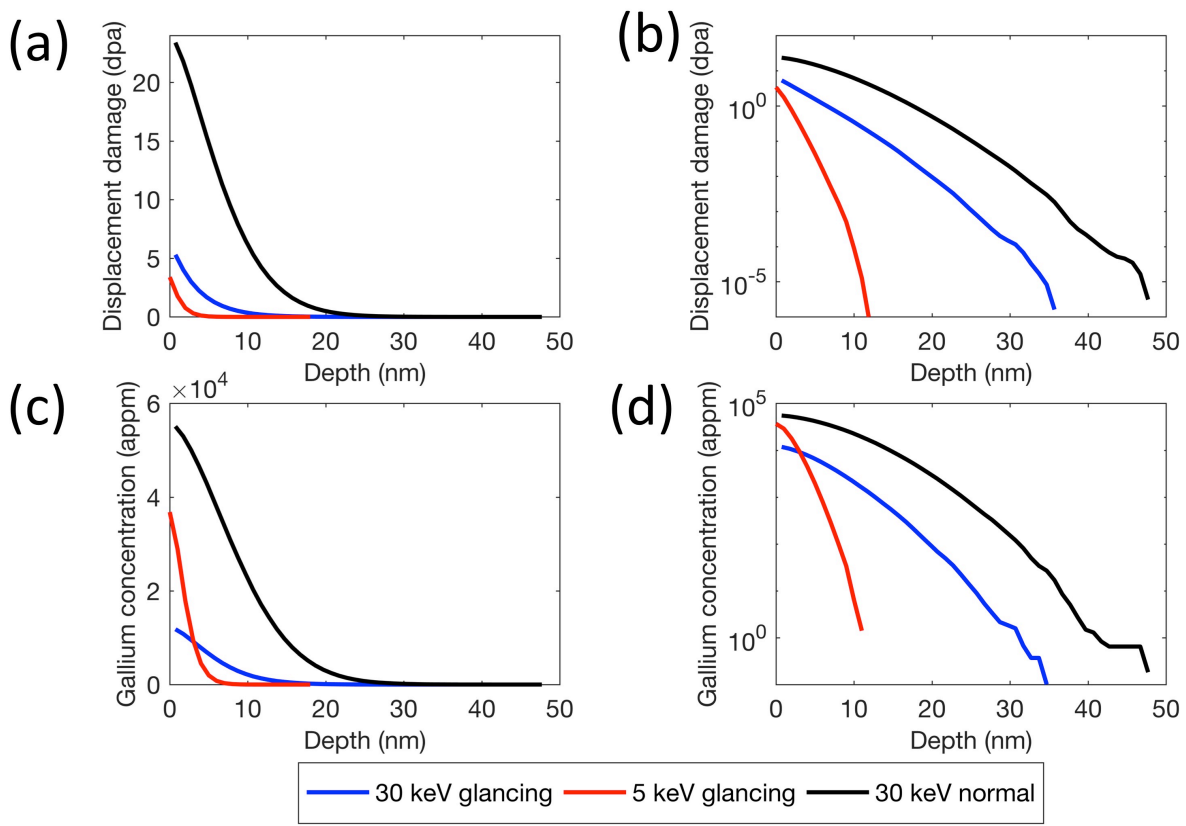

**Fig2: Predicted displacement damage and injected gallium concentration.** Profiles correspond to 30 keV normal incidence, 30 keV glancing incidence and 5 keV glancing incidence milling conditions. In all cases a fully developed milling zone with material removal greater than the damage depth was assumed. (a) and (b) Anticipated displacement damage as a function of depth on linear and log scales respectively. (c) and (d) Expected injected gallium concentration on linear and log scales respectively.

Calculations of the anticipated collision cascade damage and implanted Ga ion concentration were performed using the Stopping and Range of Ions in Matter (SRIM) code [54] ('monolayer collision – surface sputtering', 44 eV displacement energy, 3.8 eV surface energy and 3 eV binding energy for gold target [55]). Ga ions were injected at 85° and 70° from the surface normal for 30 keV glancing incidence milling and 5 keV polishing respectively. These angles were determined from the 3D crystal shape reconstructed from coherent X-ray measurements (Fig. 5 and Fig. 8). Statistics were gathered over $10^5$ ions. For 30 keV glancing incidence milling each ion is estimated to cause ~209 target displacements (~15 replacement collisions) and sputtering of ~27 Au atoms. For 5 keV polishing ~62 target displacements with ~4 replacement collisions, and a sputtering yield of 14 Au atoms per ion are predicted. The anticipated profiles of displacement damage in displacements per atom (dpa) and injected Ga-ion concentration in atomic parts per million (appm) are shown in Fig. 2. These profiles were computed taking into account the receding surface effect due to sputtering. While 30 keV glancing incidence milling produces a damage layer three times as thick as 5 keV polishing, the near-surface Ga concentration for 5 keV polishing is actually higher due to the lower sputtering yield at this energy. For comparison the profiles calculated for 30 keV normal incidence FIB milling, which produces substantially greater displacement damage and larger injected Ga concentration, are also shown.

**2.3 Bragg Coherent Diffraction Measurements**
3D resolved measurements of the lattice displacement fields in the as-made sample, after 30 keV glancing incidence FIB milling and 5 keV FIB polishing were performed using Bragg Coherent Diffraction Imaging (BCDI) at beamline 34IDC (Advanced Photon Source, Argonne National Lab, USA). Laue diffraction at beamline 34IDE (Advanced Photon Source, Argonne National Lab, USA) was used to pre-align the crystal for BCDI measurements of multiple reflections. A detailed description of the pre-alignment procedure is provided elsewhere [41]. For the as-made sample and after the 30 keV FIB milling step BCDI measurements were carried out for six crystal reflections: (200), (020), (002), (-111), (1-11), (11-1). Fig. 3 (a) shows the angular positions of the associated scattering vectors. After 5 keV polishing only three crystal reflections could be recorded ((020), (002), (-111)) before the crystal became unstable and started to rotate in the X-ray beam [56], rendering further BCDI measurements uninterpretable.

For BCDI measurements the incident X-ray beam (9 keV, ~$10^{-4}$ ΔE/E) was focussed to ~1.5 µm at the sample using KB mirrors. By positioning the micro-crystal in the KB focal plane, within the central maximum of the beam, a plane wave illumination is achieved. Diffraction patterns were recorded on a Medipix2 area detector (256 x 256 pixels, 55 µm pixel size, 16 bit image depth). For measurements of the as-made crystal a sample-to-detector distance of 1.25 m was used. For scans after FIB milling and polishing this was reduced to 1.1 m. Detector distances were determined by positioning the detector at the minimum distance required for oversampling and then increasing the distance until the diffraction patterns filled the detector matrix. A 3D coherent X-ray diffraction pattern (CXDP) was recorded from each crystal reflection by rocking through the Bragg condition, covering an angular range of -0.5° to 0.5° with respect to the reflection centre in 0.01° increments. The exposure time for each diffraction image was 0.5 s. To optimise signal-to-noise ratio, each CXDP measurement was repeated several times. These repeated scans were aligned in 3D to maximise their cross-correlation, and scans with a normalised cross-correlation-coefficient greater than 0.985 were averaged to return the CXDP for each reflection. The number of scans (noted in '') that were averaged for each reflection are: As-made crystal: (200) '12', (020) '12', (002) '12', (-111) '12', (1-11) '12', (11-1) '12'. After 30 keV FIB milling: (200) '13', (020) '10', (002) '13', (-111) '13', (1-11) '14', (11-1) '10'. After 5 keV FIB polishing: (020) '9', (002) '14', (-111) '14'.

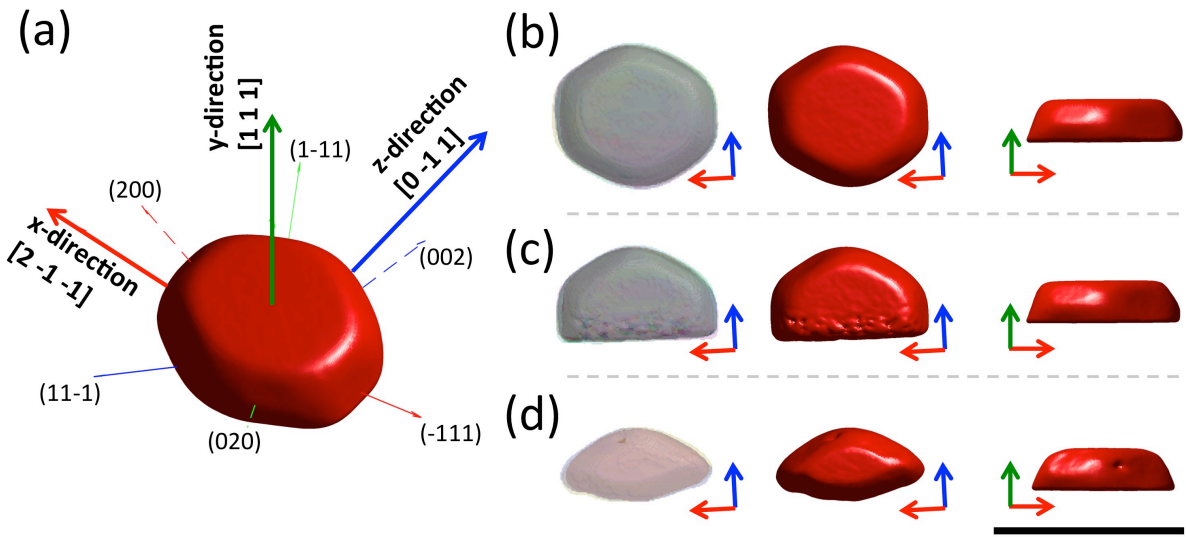

**Fig3: Reconstructed crystal shape at different milling stages.** (a) Reconstructed morphology of the as-made crystal. Superimposed are arrows indicating the crystal reflections for which CXDPs were measured (thin arrows). Also shown are the x, y and z-axes used for analysis of lattice displacements and strains, along with the corresponding real-space lattice directions (bold arrows). (b), (c) and (d) respectively show the recovered morphology of the as-made crystal, after 30 keV glancing-incidence FIB milling, and after 5 keV FIB polishing. The left column shows the superposition of semi-transparently rendered crystal morphologies recovered from different crystal reflections. The middle and right columns respectively show a top-down and side-on view of the crystal. The scalebar corresponds to 1 μm.

### 2.4 Phase Retrieval

To reconstruct the real-space electron density from a CXDP, the phase information of the diffracted wavefield must be recovered [57]. Phase retrieval was performed independently for each crystal reflection using previously published algorithms [58]. Briefly, a guided phase retrieval approach [59] with 30 random starts and 5 generations was used. For each generation 330 phase retrieval iterations were carried out (repeating pattern of 10 Error Reduction (ER) and 40 Hybrid Input Output (HIO) iterations ($\beta$ = 0.9), finishing with 30 ER iterations [60]). The reconstruction returned after each generation was the average of the final 10 ER iterations. A sharpness metric, previously shown to provide the most reliable reconstructions for strained samples [61], was used to rank the quality of reconstructions. After each generation the 3 least promising reconstructions were abandoned. After the final generation the reconstruction with the best sharpness metric was returned. Averaging over the 3 best reconstructions yielded almost identical results. Reconstructions with a greater number of iterations showed no significant further evolution of the solution.

Low spatial resolution data was used for the initial phasing generations, from which reconstructions with progressively higher spatial resolution were seeded [61]. Low resolution data was artificially generated by multiplying the 3D CXDP with a 3D Gaussian of width σ, given as a fraction of the full array size. For generations 1, 2 and 3, σ = 0.1, σ = 0.4 and σ = 0.7 was used respectively. From generation 4 onwards, full resolution data was phased. The real-space support was computed using a 3D version of the Shrinkwrap algorithm [62] and updated every 5 iterations. A 3D normalized mutual coherence function (MCF) was used to account for both longitudinal and transverse partial coherence of the illumination [58] and updated every 10 iterations. The recovered MCF had approximately Gaussian shape and magnitude >0.8, indicating an almost fully coherent illumination.

After phase retrieval in the detector coordinate frame, reconstructions were transformed to an orthogonal laboratory frame [63]. Spatial resolution was estimated by taking the derivative of electron density magnitude variation across crystal-air interfaces. The full width at half maximum (FWHM) of a Gaussian function fitted to the peak of the derivative provides an indication of the spatial resolution. For each reconstruction six line profiles were extracted (two in each spatial direction). Averaging over all reconstructions we estimate a 3D spatial resolution of ~25 nm.

X-ray propagation in the micro-crystal gives rise to an additional phase contribution since the refractive index of gold is not unity [64]. The phase change due to refraction, is given by:

$$\Delta \phi_r = 2\pi \delta \frac{l}{\lambda}, \qquad (1)$$

where $l$ is the sum of the path lengths of the incident and exit X-ray beams through the crystal to each point inside the crystal [64]. $\lambda$ the X-ray wavelength in vacuum and $\delta$ the real part of the refractive index $n$, given by $n = 1 - \delta + i\beta$. For each reconstruction the total X-ray path length associated with every voxel was computed. Then, using $\delta = 3.735 \times 10^{-5}$ [65], the refractive-index-induced 3D phase was estimated and subtracted from the reconstructed phase. Phase ramps were removed by re-centring the Fourier transform of the complex electron and phase wraps were unwrapped using the algorithm of Cusack et al. [66]. Finally the reconstructions from all measured reflections were mapped into the same coordinate system (Fig. 3) [41].

Fig. 3 (a) shows the average morphology of the as-made micro-crystal recovered from 6 measured reflections. To assess agreement of the morphology reconstructed from individual reflections, Fig. 3(b) (1st column), shows a semi-transparent rendering of the shape determined from each reflection. Clearly agreement is excellent. The recovered morphology (Fig. 3(b) (2nd column) also agrees very well with an SEM image of the crystal recorded from the same viewpoint (Fig. 1(a)). Fig. 3 (c) and (d) show the same plots for the crystal after FIB milling and polishing respectively. The morphologies recovered from different crystal reflections agree well, with only small differences at the edges of the reconstruction after FIB polishing. The top down view (2nd column of Fig. 3 (b) – (d)) clearly shows the removal of material due to the FIB processing steps. From the side views (3rd column of Fig. 3 (b) – (d)) it is clear that the crystal height remains unchanged. This confirms that material was predominantly removed from the crystal face exposed to glancing incidence FIB machining.

## 2.5 Recovery of 3D displacement and strain fields

The phase of the complex-valued electron density recovered from a specific hkl reflection, $\phi_{hkl}(\mathbf{r})$, is given by $\phi_{hkl}(\mathbf{r}) = \mathbf{q}_{hkl} \cdot \mathbf{u}(\mathbf{r})$. Here $\mathbf{u}(\mathbf{r})$ is the lattice displacement field, $\mathbf{q}_{hkl}$ the scattering vector of the reflection and $\mathbf{r}$ is the spatial coordinate. If three non-collinear reflections are measured, the system of linear equations can be solved directly to find $\mathbf{u}(\mathbf{r})$, as is the case for the measurement of the crystal after FIB polishing. If more then three reflections are measured, the system of equations is over determined. For the as-made crystal, as well as after FIB milling, six reflections were measured and a solution for $\mathbf{u}(\mathbf{r})$ was sought that minimises $\sum (\phi_{hkl}(\mathbf{r}) - \mathbf{q}_{hkl} \cdot \mathbf{u}(\mathbf{r}))^2$, where the summation is over all measured reflections [41]. The 3D-resolved lattice strain tensor, $\boldsymbol{\varepsilon}(\mathbf{r})$, can then be computed by differentiating $\mathbf{u}(\mathbf{r})$ [67].

## 3. Results and Discussion
### 3.1 As-grown micro-crystal

Fig. 4 shows the six components of the lattice strain tensor measured in the as-made micro-crystal, plotted on a virtual section through the crystal. Strains are plotted in the xyz coordinate frame shown in Fig. 3 (a) (x-axis along [2 -1 -1], y-axis along [1 1 1] and z-axis along [0 -1 1]), which is used throughout this paper.

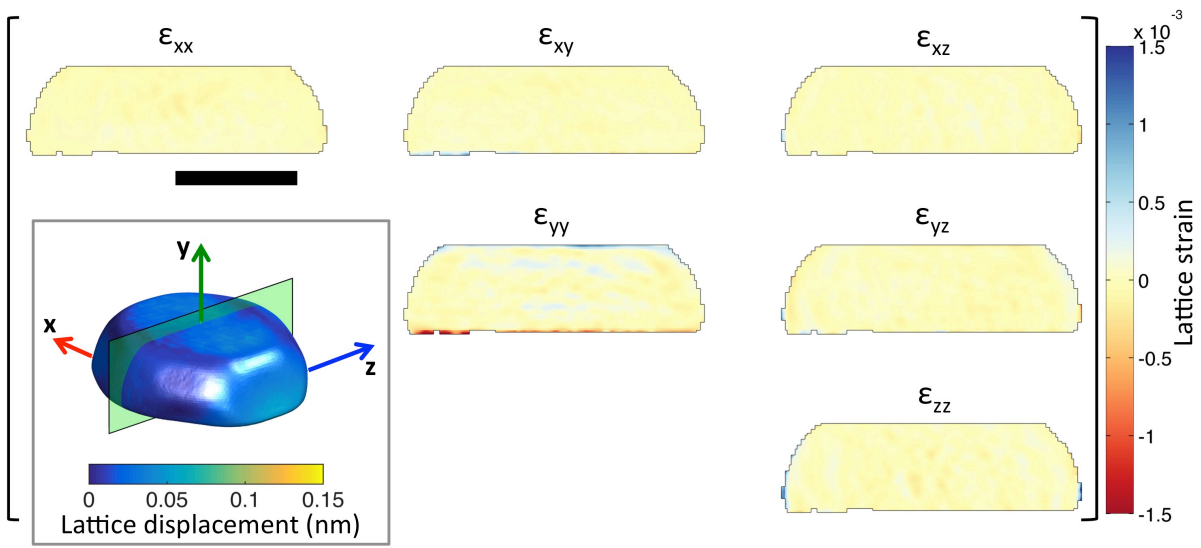

**Fig4: Strains in the as-made crystal.** All 6 components of the lattice strain tensor are shown, plotted on a section in the yz plane. The scalebar corresponds to 300 nm. Inset is a 3D rendering of the crystal coloured according to the measured lattice displacement magnitude. Superimposed are arrows indicating the directions of the x, y and z axes. The location of the plane on which strains are plotted is also shown.

Lattice displacements (Fig. 4 inset), and lattice strains in the as-made sample (Fig. 4) are small compared to the FIB-induced distortions reported below. No large phase features, such as those associated with dislocations [46], are present in the reconstructions of the as-made sample. The mean of the 6 strain components and the associated standard deviations (denoted by ±) in micro strain (x $10^{-6}$) are: $\varepsilon_{xx}$ = 3 ± 60, $\varepsilon_{yy}$ = 6 ± 209, $\varepsilon_{zz}$ = 8 ± 69, $\varepsilon_{xy}$ = 1 ± 69, $\varepsilon_{yz}$ = -6 ± 70, $\varepsilon_{xz}$ = -1 ± 44. These observations confirm that the as-made, annealed micro-crystal is practically defect free. The standard deviations of the measured strains suggest a strain uncertainty of ~2 x $10^{-4}$ in our measurements.

### 3.2 30 keV glancing incidence FIB milling

After 30 keV glancing incidence FIB milling, substantially larger lattice displacements are measured (Fig. 5 (a) inset). They are concentrated at the ion-milled surface and in particular at the top and side edges. The reconstructed 3D-resolved lattice strain tensor, plotted on a section through the crystal (Fig. 5 (b)), now shows large strains near the ion-milled surface. The $\varepsilon_{zz}$ strain component (strain normal to the ion-milled surface) is negative (lattice contraction) within a ~20 nm layer at the ion-implanted surface, followed by a ~40 nm thick layer with lattice dilatation. The strain components in the plane of the milled surface, $\varepsilon_{xx}$ and $\varepsilon_{yy}$, differ significantly: $\varepsilon_{yy}$ shows a lattice contraction near the ion-milled surface, whilst $\varepsilon_{xx}$ strains are comparatively small. The positive and negative lobes in the $\varepsilon_{yz}$ shear strain component closely resemble the shear strain pattern we observed after normal-incidence FIB imaging [46].

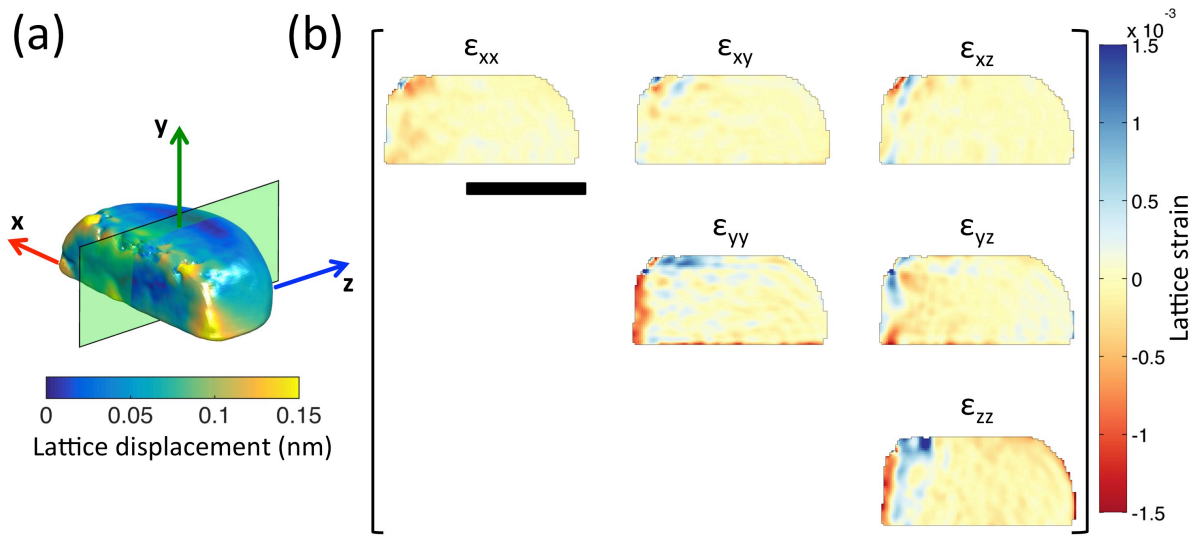

**Fig5: Lattice strains in the crystal after 30 keV glancing incidence FIB milling.** (a) Rendering of the reconstructed crystal morphology coloured according to the measured lattice displacement magnitude. Superimposed are arrows indicating the orientation of x,y and z coordinates. Also shown is the location of the plane on which strains in (b) are plotted. (b) Plots of the 6 independent components of the lattice strain tensor. The scalebar corresponds to 300 nm.

For more detailed analysis, profiles of $\varepsilon_{xx}$, $\varepsilon_{yy}$ and $\varepsilon_{zz}$, along lines in the z-direction (normal to ion-milled surface) were extracted (total of 645 line profiles on a grid of 43 x 15 positions in x and y respectively with 7 nm point spacing). The average value and standard deviation of each strain component (± error bars) are plotted as a function of distance from the ion-milled surface in Fig. 6 (a). For comparison, Fig. 6 (c) and (d) respectively show the strain profiles for 30 keV FIB imaging (50 pA, $4.2 \times 10^4$ ions/ $\mu m^2$ single imaging scan) and FIB milling (50 pA, $1.8 \times 10^8$ ions/ $\mu m^2$, 40 nm of material removed), both at normal incidence (from [46]). In all plots the strain component normal to the ion-exposed surface is plotted in red and the strain uncertainty (~$2 \times 10^{-4}$) is superimposed as a grey band.

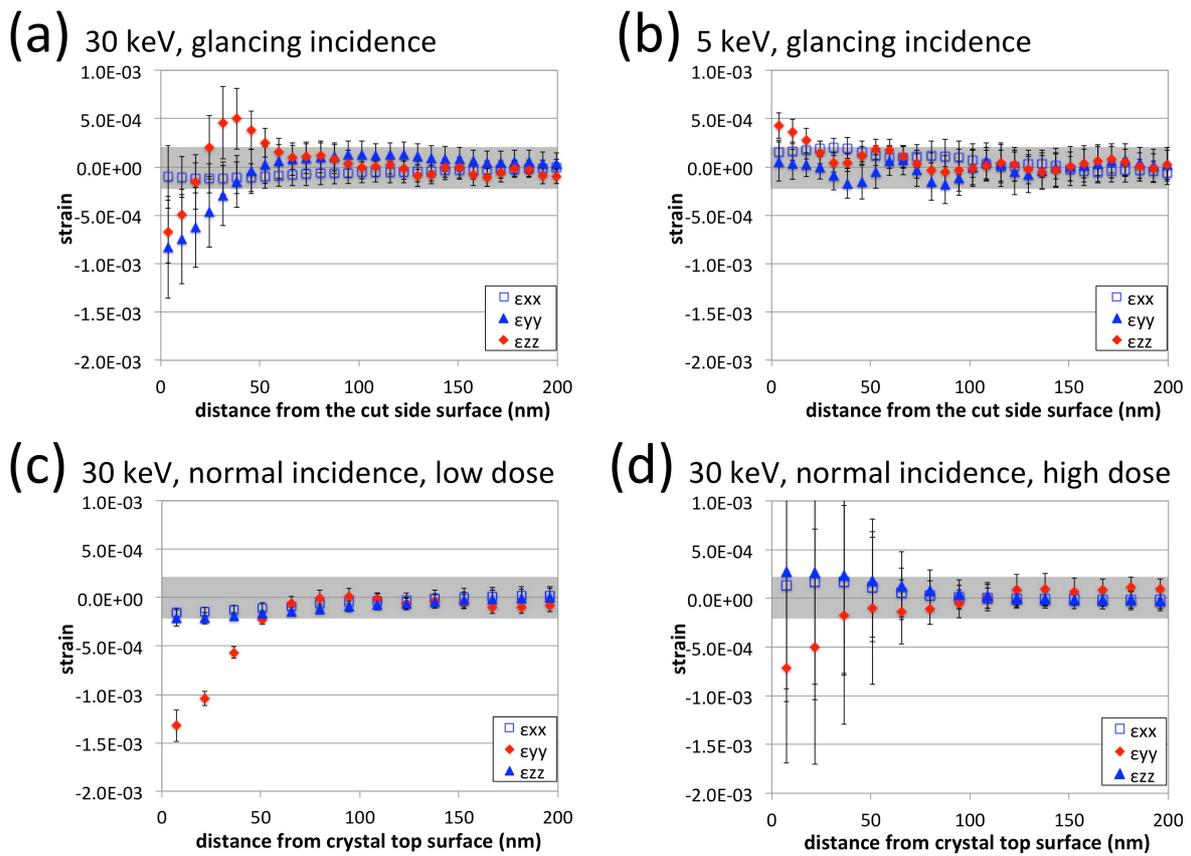

**Fig6: Comparison of FIB-induced lattice strain.** (a) and (b) respectively show strains produced by 30 keV and 5 keV glancing incidence FIB milling. Strains due to 30 keV normal incidence FIB milling at low dose ($4.2 \times 10^4$ ions/ $\mu m^2$) and high dose ($1.8 \times 10^8$ ions/ $\mu m^2$) are shown in (c) and (d) respectively. The data in (c) and (d) is described in detail in [46]. All graphs show the out-of plane direct strain (red symbols) and the two in-plane strains (blue symbols) plotted as a function of distance from the ion-milled crystal surface.

Considering the strain component normal to the ion-milled surface, several interesting observations can be made: Normal-incidence FIB only causes lattice contraction, irrespective of dose. For glancing incidence FIB milling there is lattice contraction in a ~20 nm thick layer, followed by lattice dilatation in a ~40 nm thick layer. The magnitude of lattice strains is similar in both cases, despite ~5 times greater dpa and injected Ga concentration in normal incidence FIB milling (Fig. 2). Surprisingly the largest lattice strains were measured for FIB-imaging at normal incidence (Fig. 6 (c)), and which has the lowest dpa (max. 0.025 dpa in damaged layer) and injected Ga concentration (max. 45 appm in damaged layer) [46].

For normal-incidence FIB exposure, direct strains in the plane of the ion-milled surface are small and have similar magnitude (Fig. 6 (c) and (d)). This behaviour is expected if ion-bombardment-induced defects are randomly oriented and thus give rise to a volumetric Eigenstrain, a frequently made assumption when modelling implantation damage [43,46,68,69]. For glancing incidence FIB milling a markedly different behaviour is observed (Fig. 6 (a)): The in-plane strain perpendicular to the ion-beam direction, $\varepsilon_{xx}$, is close to zero. However, the $\varepsilon_{yy}$ component, which is approximately parallel to the ion-beam, is large and negative within a ~50 nm thick surface layer. This shows that the direction of ion-implantation is important and that off-normal incidence implantation leads to an anisotropic Eigenstrain. Considering the defect dipole tensor [70,71], this suggests a preferential alignment of defects, in contrast to the usually assumed randomly-oriented defect population.

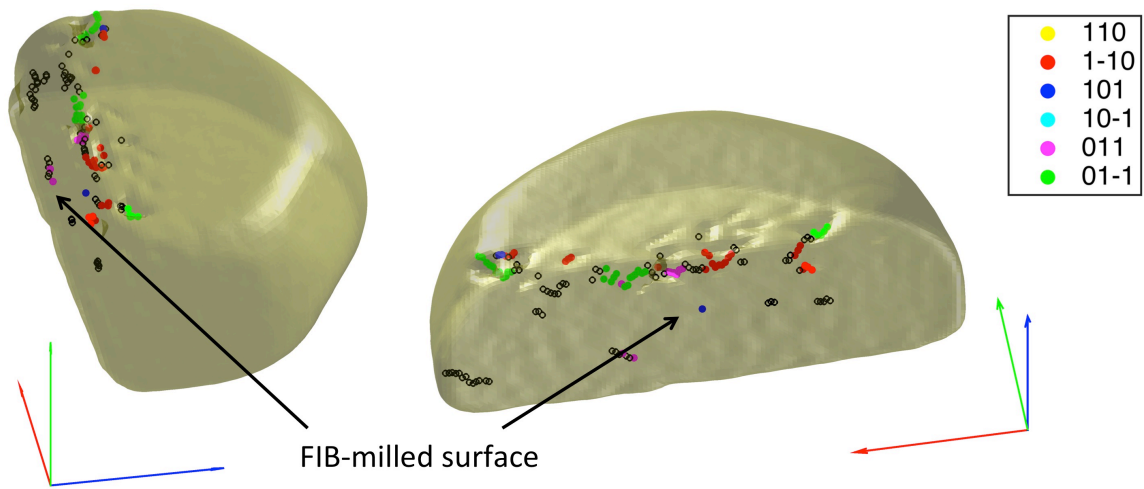

**Fig7: Defects after 30 keV glancing incidence FIB milling.** Semi-transparent rendering of the recovered crystal morphology. Superimposed are defects, coloured according to their burgers vector directions. Defects for which the burgers vector could not be unambiguously determined are shown in black. Two different viewing directions of the crystal are shown to convey the 3D arrangement of defects. The arrows in the coordinate system are plotted with a length of 300 nm.

The reconstructed crystal morphology also shows numerous "pits" near the top edge of the milled surface (Fig. 3 (c)). Large strain variations in this area can be seen in Fig. 5 (b). Together they suggest the presence of larger defects, such as dislocation lines, which manifest themselves as 'pipes of missing intensity' in BCDI measurements [46,61,72]. A dislocation with burgers vector, **b**, will only appear in the phase reconstructed from a hkl reflection if $\mathbf{q}_{hkl} \cdot \mathbf{b} \neq 0$. Thus, by considering defect visibility in several reflections, the burgers vector of specific defects can be determined. For face centred cubic (fcc) gold, dislocations with <110> burgers vector direction are expected [73]. Fig. 7 shows the dislocations that could be identified in the crystal. The FIB-milled surface incidence only shows a few small isolated defects. This is in contrast to our observations of normal incidence FIB milling, where an extended network of dislocations formed across the FIB-milled crystal face [46]. However, the large lattice strains caused by glancing incidence FIB milling (Fig. 6 (a)) indicate a high density of defects below the resolution of the present measurements.

Several dislocation loops can be seen at the top edge of the ion-milled surface (Fig. 7). These are likely to be due to the tails of the ion beam impacting the crystal at near-normal incidence. A similar effect has been reported in FIB-milled Al micro-pillars that show a dense dislocation network near the top of the as-machined pillars [38]. Indeed careful examination of TEM images from previous micro-pillar studies reveals high defect densities near the top of the undeformed samples [74,75]. In micro-mechanics samples designed for bending, the edges where such large defects would form are generally at the extremities of the crossection. During deformation these regions furthest from the neutral axis will experience the greatest strains, and hence initiation of plastic deformation is expected here. Providing additional, FIB-induced, dislocation sources in these areas could significantly modify deformation behaviour.

### 3.3 5 keV glancing incidence FIB polishing

Lattice strains in the micro-crystal after 5 keV glancing incidence FIB polishing are shown in Fig. 8. Strains near the ion-milled surface are substantially smaller than observed after the 30 keV FIB milling step. In the centre of the crystal spatial strain oscillations can be seen. These are surprising since the crystal core is expected to be largely strain free. These strain oscillations are likely to be the manifestation of an increased noise floor as only 3 reflections could be recorded after low-energy FIB polishing, compared to 6 for the as-made crystal and after high energy FIB milling. Fortunately the amplitude of strain oscillations (Fig. 8) is significantly smaller than the strains induced by FIB milling (Fig. 5) and thus does not affect our analysis.

To analyse the strains cause by low energy FIB polishing in more detail, the direct lattice strain components were extracted for lines normal to the ion-implanted surface (Fig. 6 (b)). At depths >20 nm some strain oscillations are seen as discussed above, however their magnitude is less than the estimated strain uncertainty of ~2 x $10^{-4}$ in our measurements. Within a ~20 nm thick layer of the ion-implanted surface the out-of-plane strain is positive (lattice expansion), whilst the in-plane strain components are close to zero. The near-surface magnitude of the out-of-plane strain is approximately half that measured after 30 keV glancing incidence (Fig. 6(a)) or normal incidence (Fig. 6(d)) FIB milling. The reduction of the strained layer thickness from ~60 nm for 30 keV glancing incidence milling to ~20 nm for 5 keV polishing is consistent with the calculated damage profiles (Fig. 2), which, for a given damage level, show a ~3 times thicker damage layer for 30 keV then 5 keV ion energy. In Si a much greater reduction in the damaged surface layer thickness from ~22 nm (30 keV) to 2.5 nm (5 keV) has been reported when reducing Ga-ion energy [3,30]. This suggests that the scaling of damage depth with ion energy cannot simply be based on Si. Rather the specific material system and associated damage mechanisms must be considered.

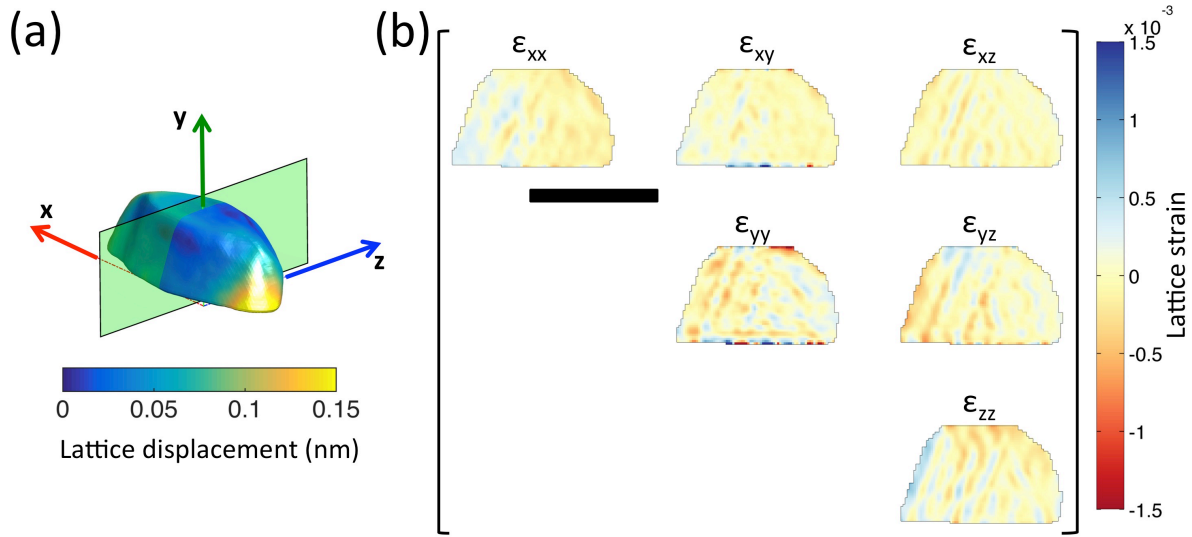

**Fig. 8: Lattice strains after 5 keV glancing incidence FIB polishing.** (a) Recovered crystal morphology coloured according to the magnitude of the lattice displacement at the crystal surface. Superimposed are the x, y and z coordinates used to plot lattice strains, as well as the plane on which lattice strains in (b) are shown. (b) Six independent components of the lattice strain tensor, plotted on a y-z section through the crystal. The scale bar is 300 nm in length.

An interesting question concerns the sign of the near-surface out-of-plane strain after 5 keV polishing. At room temperature, self-interstitials in gold are highly mobile, whilst vacancies have a higher migration energy [76–78]. The relaxation volume of a vacancy is small and negative, whilst the relaxation volume of a self-interstitial is large and positive. Thus, assuming a damage microstructure where vacancies are retained and self-interstitials migrate to the sample surface, a lattice contraction is expected. This is what we previously observed for 30 keV normal incidence FIB milling [46]. The near-surface dilatational strain after 5 keV FIB polishing may be explained by the high near-surface Ga concentration, as the relaxation volume of substitutional Ga in Au, though small, is positive [46].

By considering the amplitude and phase of the complex electron density recovered from different crystal reflections the presence of larger lattice defects can be analysed. Fig. 9 shows that the FIB-polished surface is free of extended defects. This directly demonstrates the effectiveness of low-energy FIB milling for removing defects due to previous milling steps. Surprisingly a new dislocation loop has appeared on the opposite side of the crystal, which was not exposed to ion-beam milling. Fig. 7 confirms that this dislocation loop was not present prior to the low energy FIB milling. Its exact origin remains unclear, though it might have been produced by FIB-milling-induced stress relaxation [79–81].

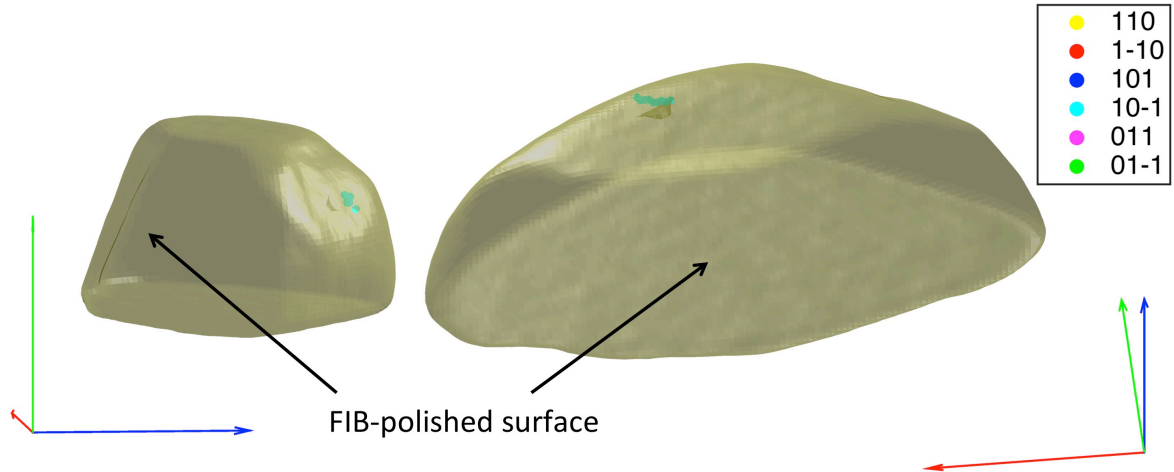

**Fig. 9: Defects after 5 keV glancing incidence FIB polishing.** Semi-transparent 3D rendering of the reconstructed crystal morphology. Superimposed are lattice defects coloured according to their burgers vector. Two views of the crystal are shown to aid visualisation of the 3D location of the dislocation loop. The axes of the coordinate system have been plotted with a length of 300 nm.

## 4. Conclusions

We have a presented a detailed experimental investigation of the residual lattice strains and defects produced by glancing-incidence focussed ion beam machining. Using multi-reflection Bragg coherent X-ray diffraction we could probe the lattice distortions inside an initially pristine gold micro-crystal after subsequent 30 keV and 5 keV milling steps. Our results show that while 30 keV glancing incidence FIB milling introduces fewer large defects than normal incidence milling, the lattice strains caused by both methods have similar magnitude and extend over 50 nm into the sample. At sample corners, where the ion beam tails impact the surface at near-normal incidence, we observe the formation of large dislocation loops, similar to those found after normal incidence FIB milling. Thus, for applications where minimising FIB-induced strains and defects is important, additional steps to remove damage caused by high-energy FIB milling must be taken. Importantly this is the case irrespective of whether normal or glancing incidence FIB milling conditions were employed. Our observations of the same crystal after 5 keV FIB polishing show that large defects and lattice strains caused by previous higher energy milling steps could be successfully removed. However, even at this lower energy, FIB-induced strains can be seen within a ~20 nm thick surface layer. Our results highlight the imperative need to carefully account for and manage FIB-induced damage. Furthermore they demonstrate that low energy ion-polishing provides an effective approach for minimising FIB-induced strains in micro-mechanics and strain microscopy samples.

**Acknowledgements**
We acknowledge funding from the European Research Council (ERC) under the European Union's Horizon 2020 research and innovation programme (grant agreement No 714697). This research used resources of the Advanced Photon Source, a U.S. Department of Energy (DOE) Office of Science User Facility operated for the DOE Office of Science by Argonne

National Laboratory under Contract No. DE-AC02-06CH11357. Use of the Center for Nanoscale Materials, an Office of Science user facility, was supported by the U.S. Department of Energy, Office of Science, Office of Basic Energy Sciences, under Contract No. DE-AC02-06CH11357. This work was in part carried out within the framework of the EUROfusion Consortium and received funding from the Euratom research and training programme 2014-2018 under grant agreement No 633053. The views and opinions expressed herein do not necessarily reflect those of the European Commission. Work performed at Brookhaven National Laboratory was supported by the US Department of Energy, Office of Basic Energy Sciences, contract DE-SC00112704.